\documentclass[journal=apchd5,manuscript=article]{achemso}

\usepackage[version=3]{mhchem} 
\usepackage[T1]{fontenc}

\author{Chengyuan Wang}
\altaffiliation{These two authors contributed equally}
\author{Ya Yu}
\altaffiliation{These two authors contributed equally}
\author{Yun Chen}
\author{Jinwen Wang}
\author{Xin Yang}
\author{Shuwei Qiu}
\author{Dong Wei}
\affiliation{MOE Key Laboratory for Nonequilibrium Synthesis and Modulation of Condensed Matter, Shaanxi Province Key Laboratory of Quantum Information and Quantum Optoelectronic Devices, School of Physics, Xi`an Jiaotong University, 710049, China}

\author{Mingtao Cao}
\affiliation{Key Laboratory of Time and Frequency Primary Standards, National Time Service Center, Chinese Academy of Sciences, Xi'an 710600, China}

\email{caomingtao1987@163.com}

\author{Hong Gao}
\email{honggao@mail.xjtu.edu.cn}
\author{Fuli Li}
\affiliation{MOE Key Laboratory for Nonequilibrium Synthesis and Modulation of Condensed Matter, Shaanxi Province Key Laboratory of Quantum Information and Quantum Optoelectronic Devices, School of Physics, Xi`an Jiaotong University, 710049, China}

 \title{Towards efficient quantum memory of orbital angular momentum qubits in cold atoms}


\begin{document}
\begin{abstract}
  The spatial modes of light, carrying a quantized amount of orbital angular momentum (OAM), is one of the excellent candidates that provides access to high-dimensional quantum states, which essentially makes it promising towards building high-dimensional quantum networks. In this paper, we report the storage and retrieval of photonic qubits encoded with OAM state in the cold atomic ensemble, achieving an average conditional fidelity above 98\% and retrieval efficiency around 65\%. The photonic OAM qubits are encoded with weak coherent states at the single-photon level and the memory is based on electromagnetically induced transparency in an elongated cold rubidium atomic ensemble. Our work constitutes an efficient node that is needed towards high dimensional and large scale quantum networks.
\end{abstract}

\section{Introduction}
Large scale quantum networks with functional quantum nodes play a pivotal role in future quantum information technologies \cite{kimble2008quantum, lvovsky2009optical, northup2014quantum}. Central to this endeavor is the large information capacity of the quantum repeaters, which can help to achieve useful communication rates in the long-distance reality \cite{collins2007multiplexed, nunn2013enhancing}. Essentially, the information capacity can be guaranteed by coding the photons in a high-dimensional space, for instance, using degree-of-freedom (DOF) of the photons, such as temporal \cite{usmani2010mapping}, spectral \cite{sinclair2014spectral}, and spatial domains \cite{lan2009multiplexed}, then the information carried by each photon could be increased significantly. In this sense, high dimension quantum memories are required to construct a large capacity quantum repeater.

Pioneering works of quantum memories with high dimension states have been demonstrated in both crystal \cite{usmani2010mapping, koroteev1999high,zhang2016experimental,kutluer2017solid} and atomic ensemble \cite{ding2014toward,parigi2015storage} by using angular modes  \cite{chrapkiewicz2017high}, spatial modes  \cite{pu2017experimental}, time-bin modes  \cite{saglamyurek2011broadband}, and OAM modes \cite{ding2014toward}. Among these proposals, the OAM mode with infinite-dimensional spanning in Hilbert space makes it an excellent candidate for high dimensional quantum memories. Since the first demonstrations of quantum memory of OAM qubits in an atomic ensemble reported \cite{ding2013single, nicolas2014quantum}, many attempts have been made to extend the merits of high dimensional storage, such as hybrid freedoms \cite{yang2018multiplexed} and high efficiency. Principally, a storage efficiency (SE) exceeding the important 50\% threshold would as well enable protocols to perform in the no-cloning regime without post-selection \cite{grosshans2001quantum} or error correction for qubit losses in linear optics quantum computation \cite{varnava2006loss}. In this endeavor, extensions towards efficient storage of high-dimensional OAM states would be crucial for building the high dimensional and large scale quantum networks. Despite the recent demonstrations of efficient quantum memories for polarization qubits \cite{vernaz2018highly,wang2019efficient,hsiao2018highly,cao2020efficient}, the highest storage and retrieval efficiencies achieved for high dimensional OAM qubits are below 30\%. The main reason is that efficient quantum memory of OAM state greatly relies on the ultra-high optical depth (OD) of the atomic ensemble, which is hard to achieve in a compressed elongated ensemble. Since the transverse size of the OAM modes is challenging to overlap with the most condensed part of the atomic ensemble, it would be applicable to win the memory efficiency by coming at the expense of lower OD.

In this paper, we experimentally realize the quantum memory (QM) for OAM qubits with the electromagnetically-induced transparency (EIT) scheme in a cold $^{87} \mathrm{Rb}$ atomic ensemble. By combining the techniques such as increasing the OD of the medium, using the $D_{1}$-line energy level, and optimizing the probe beam waveform, we obtain a SE around 65\% and fidelity above 98\% at the single-photon level. Our results may promote the high-dimensional quantum memory towards practical quantum information applications.

\section{Experimental configuration}

\begin{figure*}[!htbp]
  \centering
  \includegraphics[width=5.0in]{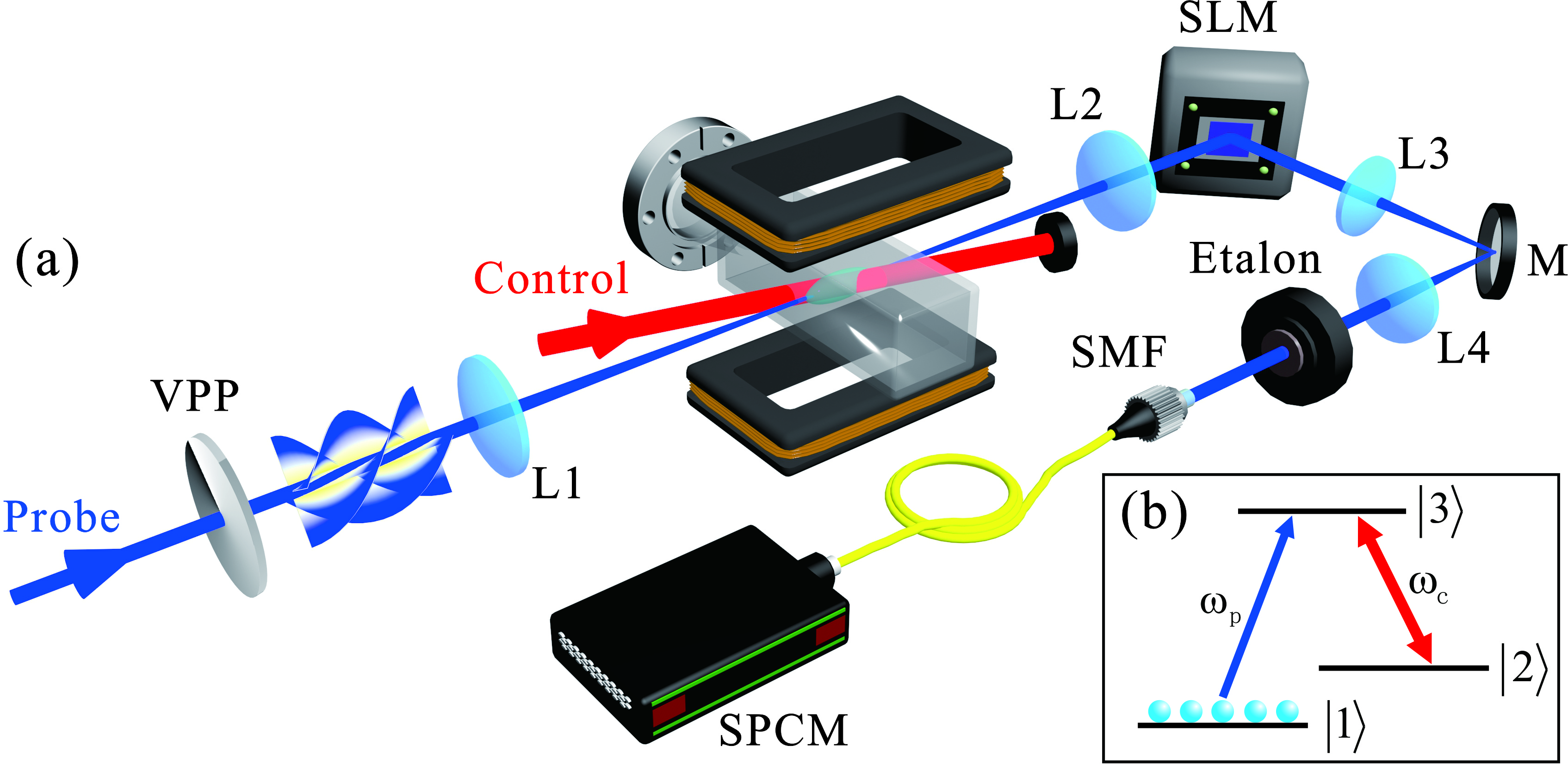}
\caption{Quantum memory for OAM qubits in a cold atomic ensemble. (a) The schematic of the experimental setup. The single-photon level probe beam is encoded with OAM via a vortex phase plate (VPP), and then it is stored in an enlongated ensemble of laser colled $^{87} \mathrm{Rb}$ atoms prepared in a two-dimensional magneto-optical trap. The retrieved OAM qubits is analysed by the projection measurement system, which composed by a spatial light modulator (SLM) and a single-mode fiber (SMF). (b) The simplified energy level structure of the experiment. $\left|1\right\rangle$ and $\left|2\right\rangle$ correspond to $\left|5 \mathrm{S}_{1 / 2}, F=1\right\rangle$ and $\left|5 \mathrm{S}_{1 / 2}, F=2\right\rangle$, which are the two hyperfine ground states of $^{87} \mathrm{Rb}$ $D_{1}$-line, and $\left|3\right\rangle$ is the excited state $\left|5 \mathrm{P}_{1 / 2}, \mathrm{F}^{\prime}=2\right\rangle$.}\label{fig:1}
\end{figure*}

To obtain an atomic cloud with large OD for storing OAM qubits, a two-dimensional (2D) magneto-optical trap (MOT) system is employed. The system composes of six counter-propagating cooling laser beams and two pairs of rectangular quadrupole magnetic coils. Each cooling beam, with 60 mW power and 3 cm beam diameter, is red detuned from $\left|5 \mathrm{S}_{1 / 2}, F=2\right\rangle \rightarrow\left|5 \mathrm{P}_{3 / 2}, \mathrm{F}^{\prime}=3\right\rangle$ by 20 MHz. The repump beam has a maximum power of 25 mW and a diameter of 3 cm. 

Each experiment period has a duration of 40 ms (25 Hz repetition rate). After an initial loading time of 30 ms, the current of the magnetic coils is increased linearly from 3 A to 12 A for the compressing MOT (C-MOT) operation, and meanwhile, the repump power is gradually decreased from 25 mW to 2 mW to employ the temporal dark-MOT technique within 7 ms. These two techniques can further increase the atomic density\cite{hsiao2014cold,jun2015cold,park2017light}. More details are given in the supplementary material. After the compressing step, all the beams and the current of the magnetic coils are turned off, and the control beam is introduced to populate most of the atoms to $\left|5 \mathrm{S}_{1 / 2}, F=1\right\rangle$ level. To decrease and eliminate the residual magnetic field, the experiment is conducted 1.2 ms after the magnetic coils are turned off. Three pairs of Helmholtz coils are utilized to further compensate the stray magnetic fields. At the end of the atom preparation stage, we obtain a OD of 220 for the probe beam transition.

After obtaining a large OD atomic ensemble, we conduct the quantum memory experiment with the EIT protocal. The probe beam will experience a large group-velocity delay, and its wave-packet is compressed inside the atomic ensemble with the presence of the control beam. When the control beam is adiabatically switched off, the probe beam is absorbed and converts into the atomic ground state coherence. The experimental setup is depicited in Fig.~\ref{fig:1}(a). The collimated probe beam can carry an OAM after passing through a vortex phase plate (VPP-m780, PRC Photonics). Lenses L1 and L2 (each has a focal length of $f=750$ mm) form a 4-f system and image the light field from the VPP to a spatial light modulator (SLM, Hamamatsu, x10468) for projective measurements. The cold atomic ensemble locates at the focal plane of L1. The control beam, with 2 mm diameter and 15 mW power, intersects with the probe beam at the atomic ensemble by an angle of $1^\circ$. Two additional home-made etalons (with 50 dB isolation ratio and 60 $\%$ transmission efficiency) are utilized to filter the scattering noise of the control beam.

To achieve a high storage efficiency, carefully choose the energy level and beam polarization is very important. Considering that for the $D_{2}$-line based EIT experiment, the control beam would off-resonantly couple to the nearby excited states, which brings about extra decay channels and decreases the storage efficiency in the high OD region\cite{hsiao2018highly}. So here we use the $D_{1}$-line of $^{87} \mathrm{Rb}$, in which the control and probe beams are on-resonant to $\left|5 \mathrm{S}_{1 / 2}, F=2\right\rangle \rightarrow\left|5 \mathrm{P}_{1 / 2}, \mathrm{F}^{\prime}=2\right\rangle$ and $\left|5 \mathrm{S}_{1 / 2}, F=1\right\rangle \rightarrow\left|5 \mathrm{P}_{1 / 2}, \mathrm{F}^{\prime}=2\right\rangle$ transitions, respectively. Since atoms are initially populated on all the Zeeman levels of $\left|5 \mathrm{S}_{1 / 2}, F=1\right\rangle$, the control beam should have the same circular polarization as the probe beam, which can fully open a transparency window to decrease the absorption loss.

It was proposed that shaping the profile of the probe pulse is helpful for increasing storage efficiency \cite{chen2013coherent,novikova2007optimal}. We tailor the temporal waveform of the probe pulse to a truncated Gaussian shape with 200 ns duration, as shown in Fig.~\ref{fig:2}(a). This can guarantee that the rear edge of the pulse is inside the atomic ensemble when the control beam is ramped off, which enables a higher storage efficiency compared with a normal Gaussian shape. After one pulse delay (200 ns) time, the control beam is turned on again to release the stored light.
 
\section{Experimental results and discussions}

\begin{figure*}[!htbp]
  \centering
  \includegraphics[width=5.5in]{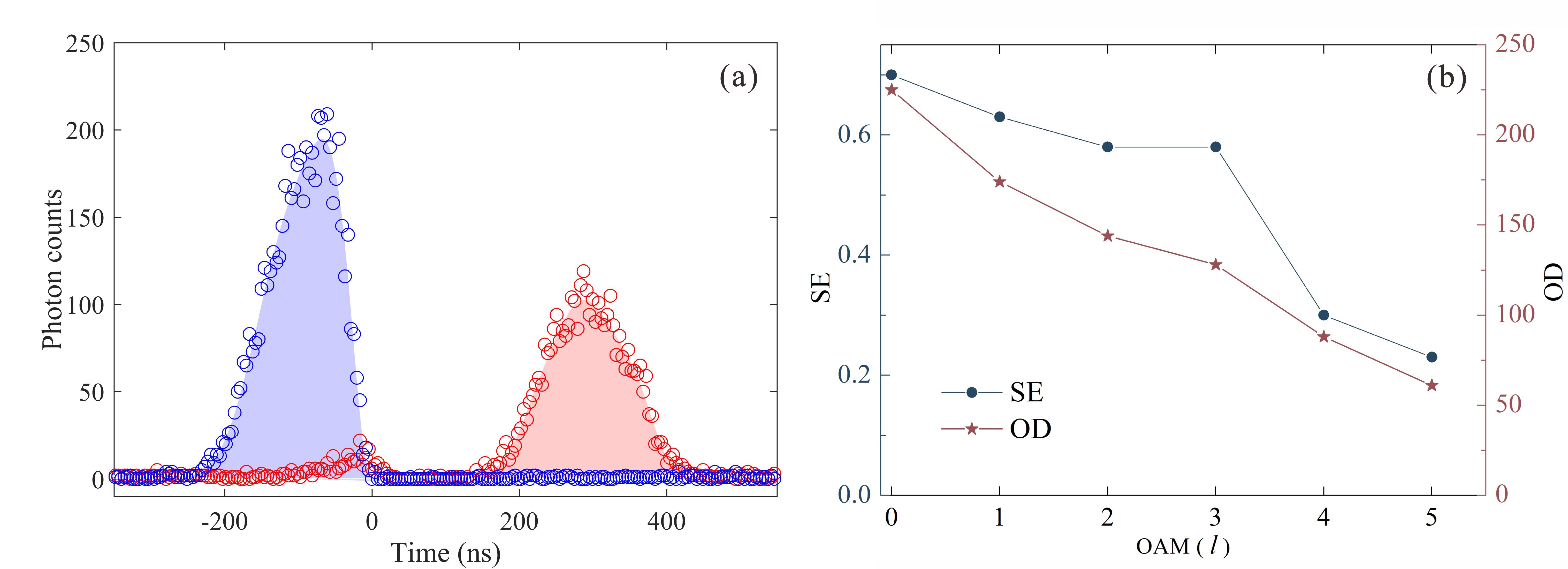}
\caption{Storage efficiency (SE) measurement for different OAM modes. (a) The temporal waveforms of the input pulse with $l=1$ (blue circle) and the retrieved signal (red circle) with one-pulse-delay storage. The photon counts are collected for 3600 s. (b) SE and optical depth (OD) against different OAM modes.}\label{fig:2}
\end{figure*}

Firstly we measure the SE of different OAM modes. Here the probe pulse is attenuated to contain a mean photon number ($\bar{n}$) of 0.5 and encoded with OAM quantum number $l$ varying from $1$ to $5$. The storage efficiency is defined as $\eta_{\text {store}}=N_{\text {out}}/ N_{\text {in}}$, where $N_{\text {in(out)}}$ represents the total recorded photon number of the input (output) probe pulse under a certain accumulation time.     
In the experiment, a single photon counting module (SPCM, Perkin-Elmer SPCM-AQR-14) is used to record $N_{\text {in(out)}}$. Figure~\ref{fig:2}(a) shows the temporal waveform of the input pulse (blue circle) and the output signal (red circle) for the case of $l=1$, with a collection time of 3600 s. Fig.~\ref{fig:2}(b) shows the one-pulse delay SE of different OAM modes. In the case of $l=1$, the SE can reach up to 65$\pm 2\%$, while for $l=5$, the SE drops to 26$\pm 2\%$. The beam waist for the OAM mode can be estimated as $\omega(z)=\sqrt{l+1}\omega_{0}(z)$, where $\omega_{0}(z)$ is the Gaussian beam waist\cite{ding2014toward}. The diameter as well as the central singular point becomes larger for a beam with higher OAM value, resulting in the decrease of the effective photon-atom interaction area (equivalently, the OD). This is the main reason for the continuous drop of SE with the increase of OAM. Further enlarging the cross-section of the cold atom, e.g. by utilizing cooling beams with larger waists, could increase the SE for high-order OAM modes.

We now turn to the storage of OAM qubits. In the following, we define the Gaussian mode ($l=0$) as $|\mathrm{G}\rangle$, OAM mode with $l=1$ as $|\mathrm{R}\rangle$. Their arbitrary superposition state $|\Psi\rangle=\alpha|\mathrm{G}\rangle+\beta e^{i\phi}|\mathrm{R}\rangle$ could be obtained by shifting and rotating the VPP \cite{wang2020generation}, here the coefficients $\alpha$ and $\beta$ denote the probability of each component and $e^{i\phi}$ is their relative phase difference. As an illustration, we perform storage for three representative OAM qubits: $|\mathrm{R}\rangle$, $|\mathrm{H}\rangle$ and $|\mathrm{A}\rangle$. The average retrieval efficiency for these three input states reaches 65\%, which is the highest SE reported so far for OAM qubits. 

Morever, to quantify the storage peformance for the OAM qubits, we reconstruct and compare the density matrices of the input and output states, which are realized by the quantum state tomography method \cite{james2005measurement}. By projecting the qubit state into three mutually unbiased bases (MUBs): $\{|\mathrm{G}\rangle, |\mathrm{R}\rangle\}$, $\{|\mathrm{H}\rangle=(|\mathrm{G}\rangle+|\mathrm{R}\rangle) / \sqrt{2},\ |\mathrm{V}\rangle=(|\mathrm{G}\rangle-|\mathrm{R}\rangle) / \sqrt{2}\}$, and $\{|\mathrm{D}\rangle=(|\mathrm{G}\rangle+i|\mathrm{R}\rangle) / \sqrt{2}, \ |\mathrm{A}\rangle=(|\mathrm{G}\rangle-i|\mathrm{R}\rangle) / \sqrt{2}\}$, the Stokes parameters $S_{i}$ can be deduced as $S_{1}=(P_{H}-P_{V})/(P_{H}+P_{V})$, $S_{2}=(P_{D}-P_{A})/(P_{D}+P_{A})$, and $S_{3}=(P_{G}-P_{R})/(P_{G}+P_{R})$, where $P_{j}$ is the total photon counts with 1200 s collection time in each projection basis. The reconstructed density matrix can thus be calculated as $\hat{\rho}=\frac{1}{2}\left(\hat{\mathrm{I}}+\sum_{i=1}^{3} S_{i} \hat{\sigma}_{i}\right)$, where $\hat{\mathrm{I}}$ is the identity matrix and $\hat{\sigma}_{i}$ are three Pauli spin operators. Then, from the measured matrices, the fidelity of the output states with the input ones could be estimated. It is defined as $\mathcal{F}_{\text {in/out}}=\operatorname{Tr}[\sqrt{\sqrt{\hat{\rho}_{out}} \hat{\rho}_{in} \sqrt{\hat{\rho}_{out}}}]^{2}$, where $\hat{\rho}_{in(out)}$ is the density matrix of the input (output) state. 

\begin{figure*}[!htbp]
  \centering
  \includegraphics[width=5.5in]{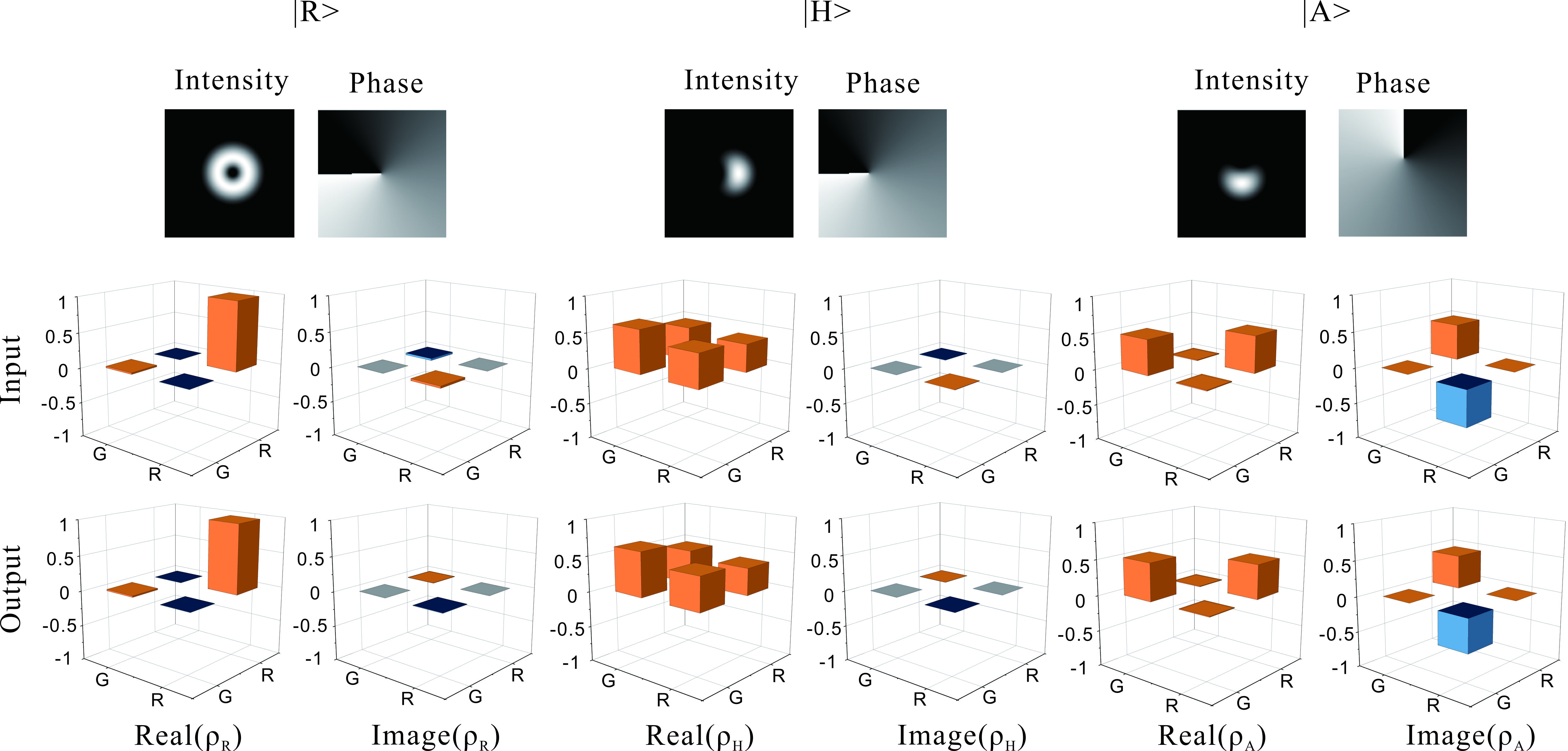}
\caption{Reconstructed density matrices of the input and output states. The input probe pulse has a mean photon number of 0.5. The data point in each projection basis is collected for 1200 s without background correction.}\label{fig:3}
\end{figure*}

The reconstructed density matrices are shown in Fig.~\ref{fig:3}. The calculated raw fidelities (with $\bar{n}$=0.5) for $|\mathrm{R}\rangle$, $|\mathrm{H}\rangle$, and $|\mathrm{A}\rangle$ are $99.4\%\pm0.2\%$, $98.9\%\pm0.5\%$, $98.3\%\pm0.3\%$, respectively. To give more faithful evidence about the quantum performance of the stored OAM qubits, we measure the fidelities for $|\mathrm{H}\rangle$ with different $\bar{n}$, which are listed in Table.~\ref{Tab:1}. To eradicate the intercept-resend attack in the context of quantum key distribution, the fidelity must be above $(N+1)/(N+2)$ with an input pulse containing $N$ photons. While for an attenuated coherent light, the probability of $N$ photons per pulse obey the Poissonian distribution $p(\bar{n}, N)=\frac{\bar{n}^{N}}{N !} e^{-\bar{n}}$. In this case the fidelity threshold can be calculated as $\mathcal{F}_{\text {coh}}(\bar{n})=\sum_{N \geq 1} (N+1)/(N+2) \frac{p(\bar{n}, N)}{1-p(\bar{n}, 0)}$ \cite{specht2011single}. As seen in Table.~\ref{Tab:1}, all the results far exceed the fidelity thresholds, indicating that the input states are able to maintain their quantum characters well after the storage and retrieval process. 

\begin{table}[htbp]
\centering
\caption{\bf Memory fidelities of the qubit state $|\mathrm{H}\rangle$ with different mean photon number per pulse without background noise correction.}
\begin{tabular}{ccccc}
\hline
Mean photon number ($\bar{n}$) & 0.1 & 0.5 & 1 & 2\\ 
\hline
Memory fidelity ($\%$) & $96.2\pm1.0$ & $98.9\pm0.5$ & $98.8\pm0.3$ & $99.1\pm0.4$\\ 
Fidelity threshold ($\%$) & $67.1$ & $68.8$ & $70.9$ & $75.0$ \\
\hline
\end{tabular}
  \label{Tab:1}
\end{table}

\section{Conclusion}

In conclusion, we demonstrate the experimental realization of high efficient EIT-based quantum memory for OAM qubits at the single-photon level. Based on a cloud of cold atoms with high OD, we achieve 65\% memory efficiency and average conditional fidelity above 98\% for the OAM qubits with a mean photon number of 0.5 per pulse. As far as we know, this is the first time to achieve a memory efficiency that exceeds 50\% for the flying qubits encoded with OAM DOF, which may have extensive application prospects in building the high dimensional and large scale quantum networks.  

\begin{acknowledgement}

The authors thank the financial support by National Natural Science Foundation of China (NSFC) (11774286, 92050103, 11534008, 12033007, and 61875205). 



\end{acknowledgement}




\bibliography{achemso-demo}








\end{document}